\documentclass[sigconf]{acmart}

\AtBeginDocument{%
  }

\copyrightyear{2024}
\acmYear{2024}
\setcopyright{acmlicensed}
\acmConference[SIGIR '24]{Proceedings of the 47th International ACM SIGIR Conference on Research and Development in Information Retrieval}{July 14--18, 2024}{Washington, DC, USA}
\acmBooktitle{Proceedings of the 47th International ACM SIGIR Conference on Research and Development in Information Retrieval (SIGIR '24), July 14--18, 2024, Washington, DC, USA}
\acmDOI{10.1145/3626772.3657736}
\acmISBN{979-8-4007-0431-4/24/07}

\usepackage{enumitem}
\usepackage{color}
\usepackage[normalem]{ulem}
\newcommand{\ie}{\emph{i.e., }}
\newcommand{\eg}{\emph{e.g., }}

\begin{document}

\title{Treatment Effect Estimation for User Interest Exploration on Recommender Systems}

\author{Jiaju Chen}
\email{cjj01@mail.ustc.edu.cn}
\affiliation{%
  \institution{University of Science and Technology of China}
  \city{Hefei}
  \country{China}
}

\author{Wenjie Wang}
\authornote{Corresponding author. This work is supported by the National Key Research and Development Program of China (2022YFB3104701), the National Natural Science Foundation of China (62272437), and the CCCD Key Lab of Ministry of Culture and Tourism.}
\email{wenjiewang96@gmail.com}
\affiliation{%
  \institution{National University of Singapore}
  \country{Singapore}
}

\author{Chongming Gao}
\email{chongming.gao@gmail.com}
\affiliation{%
  \institution{University of Science and Technology of China}
  \city{Hefei}
  \country{China}
}

\author{Peng Wu}
\email{pengwu@btbu.edu.cn}
\affiliation{%
  \institution{Beijing Technology and Business University}
  \city{Beijing}
  \country{China}
}

\author{Jianxiong Wei}
\email{weijianxiong@meituan.com}
\affiliation{%
  \institution{Meituan}
  \city{Beijing}
  \country{China}
}

\author{Qingsong Hua}
\email{huaqingsong@meituan.com}
\affiliation{%
  \institution{Meituan}
  \city{Beijing}
  \country{China}
}

\renewcommand{\shortauthors}{Jiaju Chen et al.}
\begin{abstract}
Recommender systems learn personalized user preferences from user feedback like clicks. However, user feedback is usually biased towards partially observed interests, leaving many users' hidden interests unexplored. Existing approaches typically mitigate the bias, increase recommendation diversity, or use bandit algorithms to balance exploration-exploitation trade-offs. Nevertheless, they fail to consider the potential rewards of recommending different categories of items and lack the global scheduling of allocating top-$N$ recommendations to categories, leading to suboptimal exploration. In this work, we propose an Uplift model-based Recommender (UpliftRec) framework, which regards top-$N$ recommendation as a treatment optimization problem. UpliftRec estimates the treatment effects, \ie the click-through rate (CTR) under different category exposure ratios, by using observational user feedback. UpliftRec calculates group-level treatment effects to discover users' hidden interests with high CTR rewards and leverages inverse propensity weighting to alleviate confounder bias. 
Thereafter, UpliftRec adopts a dynamic programming method to calculate the optimal treatment for overall CTR maximization. We implement UpliftRec on different backend models and conduct extensive experiments on three datasets. The empirical results validate the effectiveness of UpliftRec in discovering users' hidden interests while achieving superior recommendation accuracy. 

\end{abstract}




\maketitle

\section{Introduction}

Recommender systems are widely deployed for personalized information filtering. Technically speaking, they learn personalized user preferences from user feedback (\eg clicks and likes). However, user feedback is intrinsically biased to partially observed user interests, leaving many user interests unexplored. Consequently, recommender models learned from such biased feedback will favor partially observed user interests and aggravate the bias in the long run due to the feedback loop~\cite{10.1145/3340531.3412152,10.1145/3306618.3314288}, gradually leading to the issues such as filter bubbles and echo chambers~\cite{gao2022cirs,wenjie_fb}. Therefore, it is essential to explore users' hidden interests. 

Existing work on user interest exploration falls into three groups: 
\begin{itemize}[leftmargin=*]
\item Debiased methods alleviate the effect of biased user feedback via various reweighting or causal techniques~\cite{wang2019doubly,saito2020unbiased,chen2021autodebias,ding2022interpolative}, such as propensity weighting~\cite{saito2020unbiased} and backdoor adjustment~\cite{zhang2021causal}. However, this line of research might strengthen the minority interests while it is hard to actively discover new interests. 
\item Diversity-oriented methods~\cite{carbonell1998use,sha2016framework,stamenkovic2022choosing} aim to enrich the categories of recommended items and thus can find users' new interests. Nevertheless, such methods usually ignore the heterogeneous rewards of different item categories and probably recommend items that users dislike, leading to the sacrifice of recommendation accuracy and user satisfaction. 
\item Bandit algorithms~\cite{li2010contextual,li2016collaborative,song2022show} consider the potential rewards for enhancing the exploration-exploitation trade-off. However, recommender systems usually recommend a list of items in a session while bandit algorithms only consider the policy of a single-item recommendation, lacking global scheduling and leading to inferior rewards in a period. 
\end{itemize}

To pursue superior solutions for user interest exploration, we 
reformulate the recommendation task from a causal view. 
Generally speaking, we regard the exposure ratios of each item category in the top-$N$ recommendations as the treatment and the click-through rate (CTR) of the category as the outcome, \ie rewards. The objective of the recommender models is to maximize the potential outcome by optimizing the treatment. 
Existing recommender methods lack global scheduling for the treatments of multiple categories. They tend to over-recommend categories with more clicks, leading to an exaggeration of partially observed user interests while neglecting minority interests with potentially high rewards. 
To tackle this issue, we
have two essential considerations: 1) the precise estimation of the treatment effects, specifically the expected CTR under different category exposure ratios for each user; 2) the identification of an optimal treatment strategy to maximize the overall potential outcome for each user. 

\begin{figure}[t!]
\setlength{\abovecaptionskip}{0cm}
\setlength{\belowcaptionskip}{-0.55cm}
\centering
    \includegraphics[width=1\linewidth]{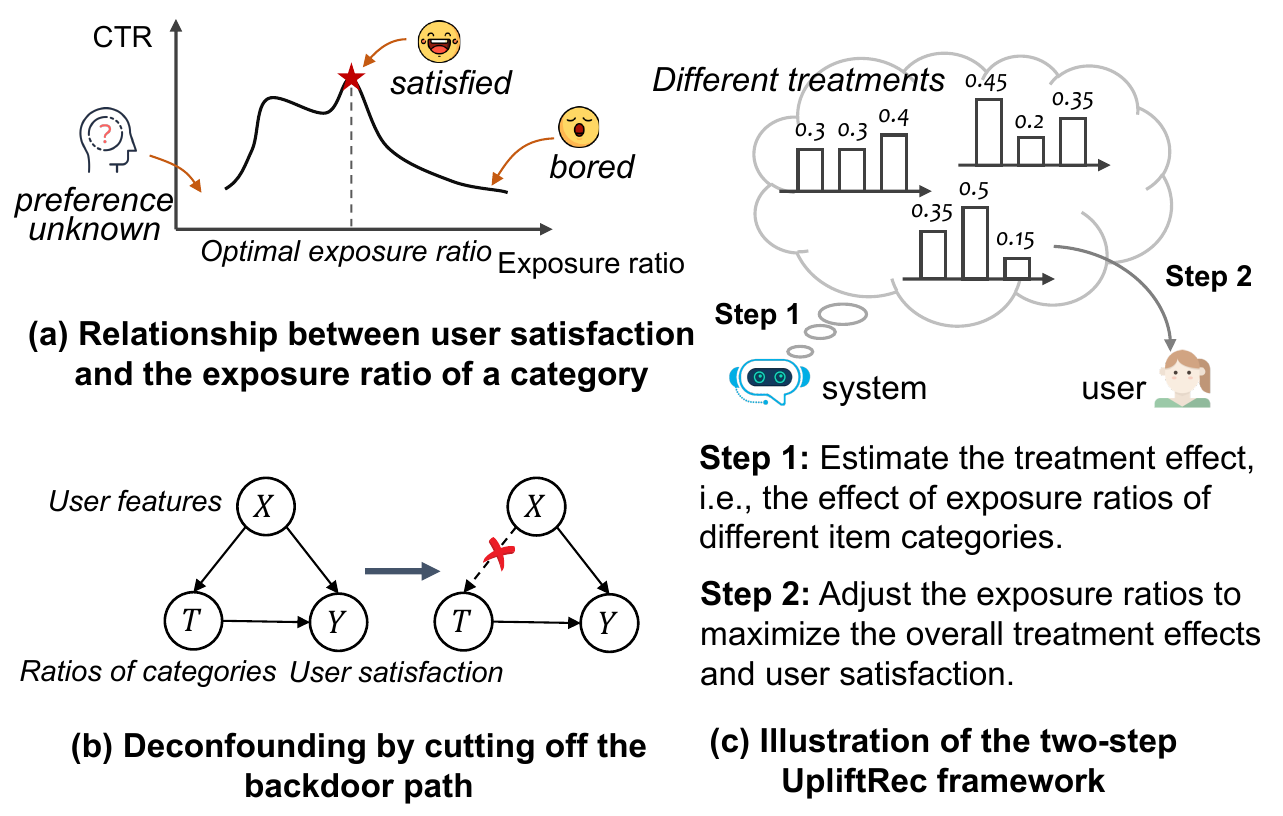}
    \caption{(a) Illustration of a user's satisfaction changing as the exposure ratio increases. (b) Illustration of the relationships of user features $X$, exposure ratios of categories $T$, and user satisfaction $Y$ from a causal view. 
    (c) UpliftRec estimates the treatment effects and schedules the optimal exposure ratios to maximize the treatment effects.}
    \label{fig:intro}
\end{figure}

Although the treatment effect estimation has been broadly studied, estimating such treatment effect for top-$N$ recommendations is non-trivial due to the following challenges. 
1) Estimating the treatment effects of recommending some minority or unexplored item categories to a user may rely on randomized controlled experiments, \ie recommending items with randomly allocated categories to this user and collecting the feedback. This is prohibitively expensive and impractical due to the potential negative impact on the user experience. 
2) Using observational user feedback for effect estimation has intrinsic biases between the treatment and outcome due to the confounding effect of users' features~\cite{zanga2022survey}, as shown in Figure \ref{fig:intro}(b). Besides, the effects of recommending unexposed item categories are missing. 
3) Worst of all, top-$N$ recommendations are essentially a problem with multivariate continuous treatments, where the exposure ratio for each category is represented by a continuous value ranging from zero to one. Hence, estimating the effects for such multivariate continuous treatments, as formulated by the Average Dose-Response Function (ADRF)~\cite{kennedy2017non,kallus2018policy}, requires a significant amount of treatment-outcome samples.

To solve these challenges, we propose an Uplift model-based Recommender (UpliftRec) framework for user interest exploration, which estimates treatment effects across multiple categories and adjusts the treatment accordingly to maximize users' total CTR, as shown in Figure \ref{fig:intro}(c). 
Specifically, 1) UpliftRec only utilizes observational user feedback for treatment effect estimation, \ie ADRF estimation. To quantify the effect on unexposed item categories (\ie users' hidden interests), UpliftRec estimates the group-level effects conditional on similar user features instead of user-level estimation, and thus leverages collaborative filtering information for users' hidden interest exploration. Intuitively, some categories unobserved by a user could be assessed by other similar users. 
Besides, 2) to mitigate the confounding biases, UpliftRec employs the inverse propensity weighting (IPW) technique for unbiased ADRF estimation. 
Lastly, 3) to efficiently utilize the observed user feedback and alleviate the high requirements for data scale, we consider discretizing continuous treatments and clustering items into condensed groups for ADRF estimation. 

Based on the estimated ADRF, we formulate the treatment selection as a linear optimization problem to maximize the overall CTR and obtain the optimal treatment assignment via dynamic programming. 
Furthermore, due to the high variance issue of ADRF, we also present a simplified effect estimation method based on the Marginal Treatment Effect Function (MTEF)~\cite{semenova2021debiased}. We conduct extensive experiments on three datasets and compare the proposed method with various competitive baselines. The empirical results show the superior performance of our method in exploring users' new interests while enhancing overall recommendation accuracy. We release the code and data at \url{https://github.com/Jiaju-Chen/UpliftRec}. 

To sum up, our contributions are threefold. 
\begin{itemize}[leftmargin=*]
    \item We examine the task of user interest exploration from a causal view and highlight the importance of estimating users' treatment effects across multiple item categories. 
    
    \item We contribute a novel UpliftRec framework, which can alleviate various challenges of estimating treatment effects, enabling reliable effect estimation and optimal treatment selection. 

    \item Extensive experiments on three datasets validate the effectiveness of the proposed method in exploring new interests while improving the CTR performance. 
    
\end{itemize}

\section{PROBLEM FORMULATION}
In this section, we first inspect the user interest exploration task. Then, we propose some basic assumptions and formulate the exploration task from a causal perspective.

\vspace{3pt}
\noindent$\bullet$ \noindent\textbf{User Interest Exploration Task.}
In this era of information explosion, recommender systems take on the responsibility of information filtering by recommending users' liked items. 
In this work, we represent users' satisfaction by the total CTR of the recommendation lists, which is a critical metric in recommendation~\cite{guo2017deepfm}. Intuitively, users favor the item categories that match their interests. However, the user feedback used for recommender training is usually biased to partial user preference, leaving many user interests unexplored. 
Besides, their interests in a certain category can be influenced by its exposure ratio, as shown in Figure \ref{fig:intro}(a). 
When the exposure ratio of a category is too low or too high, users may feel unsatisfied with the recommendations. 
As such, the key to the task lies in accurately estimating the CTR for different exposure ratios on item categories, including observed and unobserved user interests. With accurate estimation, we can determine the ideal allocation of exposure resources to each category, enhancing users' overall CTR and satisfaction. 

\vspace{3pt}
\noindent$\bullet$ \noindent\textbf{Causal View of Recommendation Process.}
Following the analysis above, we now formulate the recommendation process from a causal view. For each user $u$, we define the category exposure ratio as the treatment $T_u: t_u=[t_{u,1},t_{u,2},\dots,t_{u,C}]$, where $t_{u,c}\in[0,1]$, $\sum^C_{c=1} t_{u,c}=1$, and $C$ is the category number. And we define the potential outcome by the corresponding CTR 
$Y_u(t): y_u(t)=[y_u(t_{1}),y_u(t_{2}),\dots,y_u(t_{C})]$.  
The observed outcome $y_u(t_u)$ indicates the CTR when taking treatment $t_u$ for user $u$. 
To measure the treatment effects with varying exposure ratios, we can utilize ADRF~\cite{kennedy2017non} and MTEF~\cite{semenova2021debiased}. Formally, 
    \begin{equation} 
    \begin{aligned} \label{ADRF_def}
    \mathrm{ADRF}_u(t)&=\mathbb{E}[y_u(t)]\\
    &=[\mathbb{E}[y_u(t_{1})],\mathbb{E}[y_u(t_{2})],\dots,\mathbb{E}[y_u(t_{C})]] \\
    &\triangleq [\mathrm{ADRF}_u(t_1), ..., \mathrm{ADRF}_u(t_C)],
    \end{aligned}
    \end{equation}
    which is the expectation of CTR of each category given exposure treatment $t$ for user $u$.
    \begin{equation} 
    \begin{aligned} \label{MTEF_def}
    \mathrm{MTEF}_u(t)&=\frac{\mathbb{E}[y_u(t+\Delta t)]-\mathbb{E}[y_u(t)]}{\Delta t} \\
    &=\frac{\mathrm{ADRF}_u(t+\Delta t)-\mathrm{ADRF}_u(t)}{\Delta t},
    \end{aligned}
    \end{equation}
    which is a gradient approximation of ADRF around exposure treatment $t$, where $\Delta t$ is the interval of $t$.
 For brevity, we omit the subscript $u$ in $t_u$, $y_u(t)$, $\mathrm{ADRF}_u(t)$, and $\mathrm{MTEF}_u(t)$ in the following since all of these concepts are defined at the user level.
With ADRF, we can formulate the task of finding the appropriate exposure ratios into a constrained optimization problem:
\begin{equation}
\begin{aligned} \label{constrained_opt_problem}
&\max_{t=[t_1,t_2,\dots,t_C]}\sum^C_{c=1}t_c\cdot \mathrm{ADRF}(t_c)\\
&\begin{array}{r@{\quad}r@{}l@{\quad}l}
s.t. &\sum^C_{c=1}t_c=1; t_c\geq 0;\\
     &\sum^C_{c=1}|t_c-t_{0,c}| \leq \epsilon,\\
\end{array}
\end{aligned}
\end{equation}
where $t_0$ is a reference of category exposure ratios, possibly from the recommendation list of a backend recommender model, and $\epsilon$ controls the deviation acceptance from the reference. This constrained optimization problem can be solved by dynamic programming to calculate the best treatment (\ie the best exposure ratios across categories). Based on the best treatment, we can adjust the number of recommended items in each category. 

Additionally, we can estimate MTEF by ADRF and 
utilize MTEF to adjust the recommendation list of existing recommender models, 
which turns out to be an efficient and well-performed practice. Following~\cite{zhang2021unified}, our method is based on several assumptions:
\begin{itemize}[leftmargin=*]
    \item \textbf{Overlap.} Every subject has a non-zero probability of receiving the treatment (\ie $0 < P(t | x) < 1$) for all $t$ and $x$, where $x$ represents the observed covariate. In the context of recommender models, most systems incorporate randomness to ensure diversity, enabling that even sparse categories can have a probability of being recommended.
    \item \textbf{SUTVA.} The stable unit treatment value assumption (SUTVA) supposes that individuals do not interfere with each other. In most recommendation scenarios, users only see and give feedback to their own recommendations without the interfere from others' recommendations, satisfying the SUTVA assumption. 
    \item \textbf{Unconfoundedness.} The distribution of treatments is independent of the distribution of potential outcomes conditioned on a set of observed confounders. {Here treatments represent exposure ratios of categories by recommender systems, mainly confounded by users' features and behaviors~\cite{yu2022mdp2,schnabel2016recommendations}.} Following previous work~\cite{yu2022mdp2,ai2022lbcf,schnabel2016recommendations}, we also treat user features and interaction behaviors as the primary confounders to make the treatments independent with potential outcomes. 

\end{itemize}

\section{Method}
In this section, we describe the process of estimating ADRF for the optimization in Eq.~(\ref{constrained_opt_problem}). Subsequently, we detail how to select the optimal treatment via Eq.~(\ref{constrained_opt_problem}) using dynamic programming. Finally, we present a more streamlined method achieved through the calculation of MTEF.

\subsection{Treatment Effects Estimation}
\begin{figure}[t!]
\setlength{\abovecaptionskip}{0cm}
\setlength{\belowcaptionskip}{-0.45cm}
    \centering
    \includegraphics[width=0.7\linewidth]{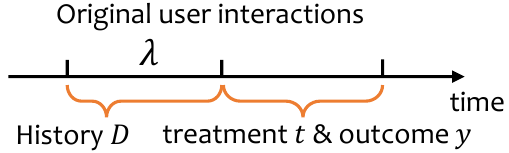}
    \caption{Illustration of generating a sample out of a real interaction trail, including history and treatment\&outcome.}
    \label{fig:sub_gen1}
\end{figure}

\vspace{3pt}
\noindent$\bullet$ \noindent\textbf{Augmented Dataset Generation.}
To estimate the users' ADRF (\ie expectation of CTR on each category), we need to observe the user feedback when applying specific treatments to each user. 
Based on observational users' interactions, we generate the samples for effect estimation in the tuple of interaction histories, treatments, and outcomes.

For each user $u$ with an interaction sequence $\hat{D}_{u}$, we generate a sample out of user $u$ with treatment and outcome data. We divide $\hat{D}_{u}$ into two parts, as is shown in Figure \ref{fig:sub_gen1}. We set the former $\lambda$ interactions as the user history part and use the latter $(1-\lambda)$ interactions to obtain the treatment and outcome, where hyper-parameter $\lambda$ ($0<\lambda<1$) is the ratio of dividing interaction trails. 
We then generate the samples with tuples of the following elements from every real user:
\begin{itemize}[leftmargin=*]
    \item History interactions $D_{u}$ are the former $\lambda$ interactions of $\hat{D}_{u}$. We use it to depict users' historical interactions before the treatment. 
    
    \item Treatments $T_{u}$ are category exposure ratios in the latter $(1-\lambda)$ interactions of $\hat{D}_{u}$. We compute $T_{u}$ from items exposed to user $u$.
    \item Potential outcomes $Y_{u}(t)$ are the CTR in the latter $(1-\lambda)$ interactions of $\hat{D}_{u}$ across all categories. The CTR of category $c$ is the number of positive items of category $c$ divided by that of all exposed items of category $c$. 
\end{itemize}



Considering that some recommendation datasets lack item category,  
before generating augmented datasets, we can cluster items using the k-means~\cite{hartigan1979algorithm} algorithm based on item embeddings. The item embeddings can be acquired by any recommender models. 

\vspace{3pt}
\noindent$\bullet$ \noindent\textbf{Confounder Debiasing.}
When inferring treatment effects from observed interaction trails, it is possible that confounding factors (such as user features denoted as $X$) influence both the treatment $T$ and the outcome $Y$. Consequently, this introduces bias into the causal relationship between $T$ and $Y$, as illustrated in Figure \ref{fig:intro}(b).To address this issue, we employ the IPW strategy, a well-known re-weighting method, to debias our estimations. Unlike methods that calculate the propensity score of a single item~\cite{schnabel2016recommendations, saito2020unbiased}, we aim to estimate the propensity score $P(t|x)$, which represents the probability of the treatment $t$ given the user feature $x$. Through IPW, we devise a debiased estimation for ADRF and demonstrate its validity as shown below, where $y$ and $t_c$ represent the outcome and treatment of a particular category $c$ for user $u$, and $x$ denotes user $u$'s feature. In our work, we represent user features using user history interactions $D_{u}$.


\begin{equation}\label{debiased ADRF} 
    \begin{split}
        \mathrm{ADRF}(t_c) &= \mathbb{E}[y(t_c)] = \sum_x P(y | t_c, x) P(x) \\
        &= \sum_x \frac{P(y, t_c| x)}{P(t_c| x)} P(x) \\
        &=  \mathbb{E}_x \left[ \frac{P(y, t_c| x)}{P(t_c | x)} \right]. 
    \end{split}
\end{equation}
Thus, by incorporating the inverse propensity $P(t_c|x)$, we effectively block the backdoor path from $T_c$ to $Y$. However, due to the sparsity of the dataset, estimating $P(t_c|x)$ using an MLP proved challenging. Therefore, we turned to estimate the propensity using a clustering approach.

We obtain the embeddings of real users and their augmented samples by training the backend model with history interactions of all of them. By computing the cosine similarity between each real user and augmented samples, we identified the top $K_p$ samples with the highest similarity and used their discretized treatment exposure ratios to calculate the propensity $P(t|x)$. Through discretization, $P(t|x)$ can be represented by a $C*(K+1)$ matrix, where $C$ is the number of categories, and $K$ is the treatment-clip number. We divided the continuous category exposure distribution into $K$ slots to handle data sparsity, and $P[c][\Tilde{t}_c]$ represents the probability of the exposure ratio on category $c$ falling into the $\Tilde{t}_c$-th slot.
\begin{equation}\label{propensity} 
    \begin{aligned}
    P[c][\Tilde{t}_c] &=\frac{\sum^{K_p}_{j=1}\mathbb{I}(Dis(t_{u_j})=\Tilde{t}_c)}{K_p}, \\ 
    & c=1,2,\dots,C; \Tilde{t}_c=0,1,\dots,K,
    \end{aligned}
\end{equation}
where $Dis$ is the discretizing function mapping continuous ratios $T$ to discrete versions $\Tilde{T}$. We set a lower threshold $v_p$ for the propensity score in case of overflow. 

\begin{figure}[t!]
\setlength{\abovecaptionskip}{0cm}
\setlength{\belowcaptionskip}{-0.35cm}
    \includegraphics[width=1.0\linewidth]{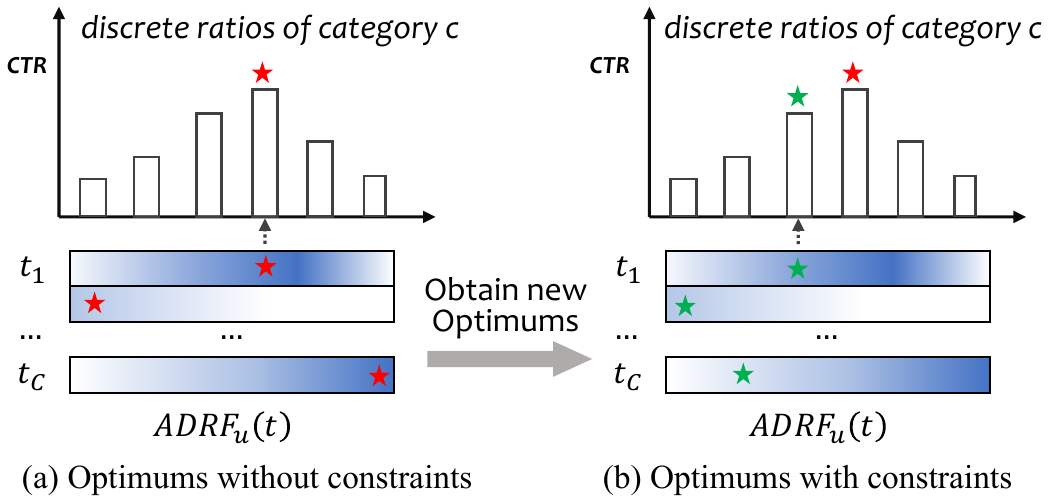}
    \caption{Optain the new optimums under the constraint --- the ratios of different categories sum to $1$.}
    \label{fig:ADRF_opt}
\end{figure}

\vspace{3pt}
\noindent$\bullet$ \noindent\textbf{ADRF Calculation.}
After preparing the augmented dataset and the propensity, we are now able to estimate users' ADRF. We derive the discrete-distribution form of ADRF for two reasons: 1) Utilizing a dynamic programming algorithm enables us to efficiently determine the best exposure distribution in linear time. 2) The optimal distribution must be discrete, considering we can only recommend integer numbers of items in the final recommendation list.

Therefore, we also represent ADRF in a matrix form with dimensions of $C\times (K+1)$, where $C$ is the category number, and $K$ is the treatment-clip number. We approach the optimization process as a knapsack problem, where the knapsack space has $K$ slots, and the value of each item is the CTR. As illustrated in Figure \ref{fig:ADRF_opt}, each row of ADRF represents the CTR of category $c$ when exposed to different discrete ratios. In the discretized scenario, we convert the continuous constraint $\sum^C_{c=1}t_c=1$ into the discrete constraint $\sum^C_{c=1}\Tilde{t}_c=K$, where $\Tilde{t}_c$ represents the discretized treatment. Our goal is to optimize the total CTR (the sum of CTR of all items in the top-$N$ list) under the discrete constraint.

We retrieve the same embedding information as the propensity calculation process. By fetching the top $K_s$ samples with the largest cosine similarity, we calculate $\mathrm{ADRF}[c][k]$ using their average $y_{u_j}$ weighted by their inverse propensity $P(t_c|x_{u_j})^\gamma$. Specifically, 
\begin{equation}\label{ADRF} 
    \begin{aligned}
     \mathrm{ADRF}[c][\Tilde{t}_c] &= \frac{\sum^{K_s}_{j=1}\mathbb{I}(Dis(t_{u_j})=\Tilde{t}_c) * y_{u_j} / P[c][\Tilde{t}_c]^\gamma }{\sum^{K_s}_{j=1}\mathbb{I}(Dis(t_{u_j})=\Tilde{t}_c)}, \\
     & c=1,2,\dots,C; \Tilde{t}_c=0,1,\dots,K,
     \end{aligned}
\end{equation}
where $t_{u_j}, y_{u_j}$ represent the treatment and outcome of sample $u_j$, $P$ is the propensity matrix, $\gamma$ is a hyper-parameter to adjust the extent of debiasing. We set the first column $\mathrm{ADRF}[:,0]$ as 0 since if a category is not exposed, its CTR is 0. Any other places without treatment-outcome data are padded with a null value $v_a$.

\subsection{Fetching the Best Treatment}

In contrast to other algorithms that primarily focus on point-wise accuracy, we propose that top-$N$ recommendation is essentially a resource allocation problem. After computing the ADRF of each real user, we transform the optimization problem into a knapsack problem using dynamic programming. We introduce a new matrix $f$ with the same shape as ADRF, where $f[c][k]$ represents the maximal overall CTR achieved by allocating $k$ slots of exposure resources to the first $c$ categories. The state transition equation is:
\begin{equation}\label{state_transition_equation} 
    \begin{aligned}
    f[c][k] &= \max_{|t_{0,c}-j|<\epsilon}(f[c-1][k], f[c-1][k-j]\\ 
    &+j*\mathrm{ADRF}[c][j]), c=1,2,\dots,C; k=0,1,\dots,K,
\end{aligned}
\end{equation}
where $t_0$ represents the category exposure distribution of the backend recommendation list, and $\epsilon$ is the allowable deviation from the distribution of the backend recommendation list. The value of $f[C][K]$ represents the maximal overall CTR that we aim to achieve, and by tracing its path, we can identify the optimal treatment, enabling us to re-allocate the recommendation list accordingly. We refer to this method as UpliftRec-ADRF.
An advantage of dynamic programming algorithms is it takes polynomial time to solve the problem. In this case, the time complexity is $O(CK\epsilon)$, where $C$ is the category number, $K$ is the treatment-clip number, and $\epsilon$ is the allowable deviation.

\subsection{MTEF Approximation}
\label{sec:MTEF}
Unfortunately, the ADRF solution faces a significant issue of high variance due to data sparsity. To address this concern, we propose a simplified version of ADRF known as MTEF estimation. MTEF aims to calculate a gradient approximation to mitigate the impact of sparse data. We consider that the farther away $t$ is from the backend distribution $t_0$, the higher the variance the model will encounter when estimating $\mathrm{ADRF}(t)$. To offer a slight tuning on the backend recommendation list, we use the treatment effects near $t_0$. $\mathrm{MTEF}(t_0)$ represents the discrete form of the gradient $\frac{\partial{y(t_0)}}{\partial t}$, which indicates whether increasing exposure boosts the potential CTR on certain categories near $t_0$.
\begin{equation}\label{MTEF_definition} 
    \mathrm{MTEF}(t)=\frac{\mathbb{E}[y(t+\Delta t)]-\mathbb{E}[y(t)]}{\Delta t},
\end{equation}
where $\Delta t$ is the step. The null values are filled by $v_m$. Notably, each user has unique ADRF and MTEF. For the sake of brevity, we omit the subscripts.

After calculating MTEF, we add it to scores provided by the backend model to obtain the final score.
\begin{equation}\label{rerank} 
    s_{u,i} = s_{0,u,i} + \alpha*\mathrm{MTEF}(t_0)[c_i],
\end{equation}
where $s_{0,u,i}$ is the backend score for (u, i), $\alpha$ is a hyper-parameter to adjust the influence of MTEF, and $c_i$ denotes the category item $i$ belongs to. We name this method UpliftRec-MTEF.


\section{Experiments}
In this section, we conduct extensive experiments to answer the following research questions: 
\begin{itemize}[leftmargin=*]
    \item \textbf{RQ1:} How does our method perform compared to SOTA debiased, diversity-based, bandit methods?
    \item \textbf{RQ2:} What is the impact of the IPW module on the performance of our method?
    \item \textbf{RQ3:} What distinguishes UpliftRec-MTEF from UpliftRec-ADRF?
\end{itemize}

\subsection{Experimental Settings}

\begin{table}[t]
\setlength{\abovecaptionskip}{-0.15cm}
\setlength{\belowcaptionskip}{0cm}
\caption{Statistics of the three datasets. "int." denotes "interactions". "TP." denotes "Training Positive". "TN." denotes "Training Negative".}
\begin{center}
\setlength{\tabcolsep}{2mm}{
\resizebox{\columnwidth}{!}{
\begin{tabular}{cccccc}
\toprule
\textbf{Dataset}        & \textbf{\#User} & \textbf{\#Item} & \textbf{\#TP. int.} & \textbf{\#TN. int.} & \textbf{Density}  \\ \midrule
\textbf{Yahoo!R3}       &  15.4K  &  1.0K  &   125.1K   & 167.9K &  1.9\%     \\ 
\textbf{Coat}           &  290  &  300  &    1.9K   & 5.1K & 8.0\%         \\ 
\textbf{KuaiRec}        &  7.1K  &  10.7K &   936.5K  & 11.6M & 16.5\%        \\ \bottomrule
\end{tabular}
}}
\end{center}

\label{tab:statistics}
\end{table}

\subsubsection{\textbf{Datasets}}
We conduct experiments on three real-world unbiased datasets: 1) Yahoo!R3 \cite{Marlin2009Collaborative}, 2) Coat \cite{rs_treatment}, and 3) KuaiRec \cite{gao2022kuairec}. These datasets pertain to music, coat, and short-video scenarios, respectively. Each dataset comprises two components: a biased part, collected from typical user interactions in the real world, and an unbiased part, gathered through a randomized controlled trial (RCT) that is free from system-induced biases. Table \ref{tab:statistics} presents the statistics for these datasets.
In our experiments, we utilize the biased part to train the model and partition the unbiased part for validation (50\%) and testing (50\%). For Yahoo!R3 and Coat datasets, we consider only ratings $\geq 4$ as positive examples. For KuaiRec, positive interactions are those with at least twice the watching time ratio, indicating that a user has watched a video at least twice.
We incorporate the unbiased part in our experiments to evaluate how effectively the models can explore users' latent interests, which may not be evident during the collection of typical user interactions. 

\subsubsection{\textbf{Baseline}}
We compare our method with several competitive methods including basic backend-candidate models, debiased methods, diversity-oriented methods and bandit algorithms: 
\begin{itemize}[leftmargin=*]
    \item \textbf{MF~\cite{koren2009matrix}.} It is as a widely used benchmark model in recommendation systems, owing to its simplicity and effectiveness. 
    \item \textbf{FM~\cite{rendle2012factorization}.} It is the most representative feature-based recommender model. {We selected features based on FM’s performance. We used ‘ID’ and ‘jacket type’ in Coat, ‘ID’ in Yahoo!R3, and ‘ID’ with categories in KuaiRec.}
    \item \textbf{LightGCN~\cite{he2020lightgcn}.} It is a simplified and effective GCN-based benchmark model.
    \item \textbf{IPS~\cite{schnabel2016recommendations}.} It is a  conventional inverse propensity weighting method. It tries to capture true user preference from biased data by re-weighting training samples. 
    \item \textbf{BC Loss~\cite{zhang2022incorporating}.} It is a debiased method which incorporates adjustments based on popularity into a contrastive loss.
    \item \textbf{iDCF~\cite{zhang2023debiasing}.} It is a debiased method utilizing proxy variables to infer the unmeasured confounder. 
    \item \textbf{MMR~\cite{carbonell1998use}.} It is a common reranking method that weighs between relevance and maximum list distance.
    \item \textbf{PMF-$\alpha$-$\beta$~\cite{sha2016framework}.} It balances relevance and diversity via reranking.
    \item \textbf{LinUCB~\cite{li2010contextual}.} It is an exploration and exploitation technique, continuously exploring while minimizing each arm’s variance.
    \item \textbf{HCB~\cite{song2022show}.} It is a hierarchical bandit framework for entire space user interest exploration.
\end{itemize}

\subsubsection{\textbf{Hyper-parameter Settings}}
For all the methods, the maximum embedding size is set to 512. We exclusively utilize positive samples for training these methods, and the negative sampling ratio is set to be either 4 or 24. We search the ADRF-similarity number $K_s$ and the propensity-similarity number $K_p$ from 10 to the number of samples with intervals of 3x. We tune the treatment-clip number $K$ in \{0.2, 0.4, 0.6, 0.8, 1\} of the length of recommendation list $N$. We tune the category-clustering number $C$ in \{2, 3, 5, 10, 15\}. For UpliftRec-MTEF, we tune the null value $v_m$ in [0, 0.5] with a step of 0.05 and the weight $\alpha$ in [0.05, 0.45] with a step of 0.05. For UpliftRec-ADRF, we tune the null value $v_a$ in \{0.01, 0.1\} and the allowable deviation $\epsilon$ in \{0, 1, 2\}. We generate a single sample for each real user with a division ratio of $\lambda$ set at 0.5.

\begin{table*}[t!]
\setlength{\abovecaptionskip}{0.2cm}
\setlength{\belowcaptionskip}{0cm}
\caption{Performance comparison between the baselines and UpliftRec-MTEF on the three datasets. The best results are highlighted in bold while the second-best ones are underlined. * implies the improvements over the best baseline are statistically significant (p-value < 0.05) under one-sample t-tests.}
    \centering
    \begin{center}
    \setlength{\tabcolsep}{3mm}{
    \resizebox{\textwidth}{!}{
    \begin{tabular}{l|ccc|cccc|cccc}
    
    \toprule
        \textbf{Dataset} & \multicolumn{3}{c|}{\textbf{Yahoo!R3}} & \multicolumn{4}{c|}{\textbf{Coat}} & \multicolumn{4}{c}{\textbf{KuaiRec}} \\ 
        \textbf{Metric} & \textbf{R@10} & \textbf{N@10} & \textbf{RUP@10} & \textbf{R@10} & \textbf{N@10} & \textbf{RUE@10} & \textbf{RUP@10} & \textbf{R@10} & \textbf{N@10} & \textbf{RUE@10} & \textbf{RUP@10} \\ \midrule
        \textbf{Random} & 0.0093 & 0.0044 & 0.0110 & 0.0307 & 0.0212 & 0.0321 & 0.0327 & 0.0009 & 0.0066 & 0.0010 & 0.0009 \\ 
        \textbf{MF} & 0.0621 & 0.0338 & 0.0735 & 0.0705 & 0.0397 & 0.0636 & 0.0751 & 0.0870 & 0.2741 & 0.0949 & 0.0875 \\ 
        \textbf{FM} & 0.0532 & 0.0267 & 0.0630 & \underline{0.0875} & \underline{0.0416} & \underline{0.0836} & \underline{0.0932} & 0.0468 & 0.1629 & 0.0498 & 0.0470 \\ 
        \textbf{LightGCN} & 0.0614 & 0.0299 & 0.0727 & 0.0796 & 0.0374 & 0.0777 & 0.0848 & 0.0797 & 0.2417 & 0.0877 & 0.0802 \\ 
        \textbf{IPS} & 0.0606 & 0.0325 & 0.0717 & 0.0686 & 0.0309 & 0.0734 & 0.0730 & 0.0886 & 0.2666 & 0.0983 & 0.0891 \\ 
        \textbf{BC Loss} & \underline{0.0694} & \underline{0.0360} & \underline{0.0822} & 0.0819 & 0.0357 & 0.0702 & 0.0873 & 0.0487 & 0.1465 & 0.0464 & 0.0490 \\ 
        \textbf{iDCF} & 0.0675 & 0.0335 & 0.0799 & 0.0755 & 0.0336 & 0.0692 & 0.0804 & {0.0346} & 0.0346 & 0.038 & {0.0348} \\
        \textbf{MMR} & 0.0557 & 0.0314 & 0.0660 & 0.0821 & 0.0380 & 0.0730 & 0.0874 & 0.0784 & 0.2616 & 0.0864 & 0.0789 \\ 
        \textbf{PMF-$\alpha$-$\beta$} & 0.0567 & 0.0301 & 0.0671 & 0.0385 & 0.0214 & 0.0398 & 0.0410  & 0.0787 & 0.2550 & 0.0869 & 0.0792   \\         
        \textbf{LinUCB} & 0.0610 & 0.0332 & 0.0723 & 0.0571 & 0.0315 & 0.0712 & 0.0608 & \underline{0.0897} & \underline{0.2758} & \underline{0.0986} & \underline{0.0901} \\ 
        \textbf{HCB} & 0.0347 & 0.0168 & 0.0411 & 0.0683 & 0.0425 & 0.0791 & 0.0728 & 0.0777 & 0.2630 & 0.0862 & 0.0781 \\ 
        \textbf{UpliftRec-MTEF} & \textbf{0.0715*} & \textbf{0.0363*} & \textbf{0.0847*} & \textbf{0.1027*} & \textbf{0.0504*} & \textbf{0.0984*} & \textbf{0.1093*} & \textbf{0.0914*} & \textbf{0.3169*} & \textbf{0.0999*} & \textbf{0.0920*} \\ \bottomrule
    \end{tabular}
    }}
    \end{center}

\label{tab:overall_per}
\end{table*}

\subsubsection{\textbf{Evaluation Metrics}}
To assess the model's performance, we employ two widely-used accuracy metrics, Recall@K (R@K) and NDCG@K (N@K)~\cite{jarvelin2002cumulated}. Additionally, to assess the model's ability to explore user hidden interests, we introduce two other metrics:

\begin{itemize}[leftmargin=*]
\item \textbf{RUE@K (UnExpected Recall@K).} This metric measures the recall of unexpected items in the top-K list, where an item is considered unexpected if its category is not among the user's top 3 categories for interaction history. Please note that Yahoo!R3 does not have this metric due to the absence of category labels.
\item \textbf{RUP@K (UnPopular Recall@K).} This metric evaluates the recall of unpopular items in the top-K list, where an item is labeled as unpopular if its interaction count in the training set falls within the bottom ninety percent of all items.
\end{itemize}

We also introduce the concept of serendipity, which refers to accidentally discovering users' hidden interests. Accordingly, we categorize the four metrics into two groups: accuracy (R@K, NDCG@K) and serendipity (RUE@K, RUP@K). {We employ recall for fine-tuning UpliftRec and most baselines, as users are interested in encountering more positive items within the top-$N$ list.
For diversity-oriented baselines, we prioritize the highest diversity with the costs of a maximal 10\% reduction of recall. }

\subsection{Overall Performance (RQ1)}
We train baselines and UpliftRec-MTEF on the three datasets and results are reported in Table \ref{tab:overall_per}. We omit more results of @20 with similar trends due to space limitations. We choose the method with the best performance among MF, FM, and LightGCN as the backend model for MMR, PMF-$\alpha$-$\beta$, LinUCB, HCB, and UpliftRec. To be specific, we choose MF for Yahoo!R3, KuaiRec, and FM for Coat. From Table \ref{tab:overall_per}, we have the following observations:

\begin{itemize}[leftmargin=*]
    \item UpliftRec-MTEF consistently outperforms other baselines on the three datasets. On R@10 and N@10, UpliftRec-MTEF yields superior performance than other baselines, which validates its strong ability to accurately predict recommendations. More importantly, UpliftRec-MTEF also performs well on RUE@10 and RUP@10, which reflects a model's capacity in exploring each user' hidden interests in fields he/she or the majority of users have not explored. The higher the value of RUE@10, the greater the likelihood for users to receive items from categories they have rarely encountered before. Similarly, a larger RUP@10 indicates increased opportunities for users to discover and receive unpopular items. These results highlight the effectiveness of UpliftRec-MTEF in mitigating exposure bias and popularity bias within recommender systems.
    \item In basic backend-candidate models, MF outperforms the other two methods (on Yahoo!R3 and KuaiRec) due to its simplicity and generalization ability. FM excels on Coat by leveraging item features in scarce data, but it performs poorly on KuaiRec. LightGCN exhibits average results against the others.
    \item As to debiased methods, IPS, BC Loss, and iDCF all stem from MF. BC Loss excels over MF on datasets like Yahoo!R3 and Coat, characterized by significant popularity bias. However, BC Loss exhibits poorer performance on KuaiRec, where primary biases are exposure and position bias. Similarly, iDCF outperforms MF on Yahoo!R3 and Coat owing to confounder learning. Nevertheless, iDCF falters on KuaiRec, where full exposure renders the interacting preference learned by iVAE~\cite{khemakhem2020variational} irrelevant. Conversely, IPS performs even worse than MF on Yahoo!R3 and Coat, primarily due to the high variance issue associated with the method.
    \item For diversity-oriented methods like MMR and PMF-$\alpha$-$\beta$, we notice that RUE@10 and RUP@10 do not increase with diversity, suggesting that diversity-oriented methods might not effectively explore users' latent interests.
    \item  Regarding bandit algorithms, LinUCB performs second-best on KuaiRec, demonstrating the efficacy of the exploration-exploitation method in short-video recommendations. However, both LinUCB and HCB perform poorly in other cases, highlighting the detrimental effects of aimless exploitation.
\end{itemize}

\subsection{In-depth Analysis (RQ2 \& RQ3)} 
To further explore our method, we design a bunch of experiments, including ablation studies, comparisons between Uplift-ADRF and Uplift-MTEF, robustness analysis to various backend models, case studies, hyper-parameter analysis and comparisons with methods with unbiased data.

\begin{figure}[tb]
\setlength{\abovecaptionskip}{0cm}
\setlength{\belowcaptionskip}{-0.35cm}
\centering
\includegraphics[scale=0.43]{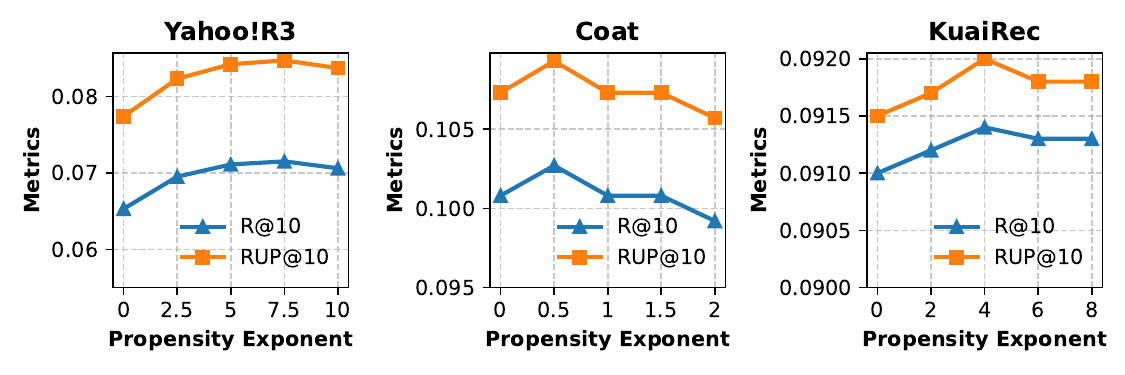}
\caption{Performance of UpliftRec-MTEF as changing propensity exponent $\gamma$ on Yahoo!R3, Coat, and KuaiRec.}
\label{fig:abs}
\end{figure}

\begin{table*}[t]
\setlength{\abovecaptionskip}{-0.1cm}
\setlength{\belowcaptionskip}{0cm}
\caption{Improvements compared to the different backend models on the Yahoo!R3 and Coat. The best results are in bold.}
    \centering
    \begin{center}
    \setlength{\tabcolsep}{6mm}{
    \resizebox{\textwidth}{!}{
    \begin{tabular}{l|ccc|cccc}
    \toprule
        \textbf{Dataset} & \multicolumn{3}{c|}{\textbf{Yahoo!R3}} & \multicolumn{4}{c}{\textbf{Coat}} \\ 
        \textbf{Metric} & \textbf{R@10} & \textbf{N@10} & \textbf{RUP@10}  & \textbf{R@10} & \textbf{N@10} & \textbf{RUE@10} & \textbf{RUP@10}  \\ \midrule
        \textbf{MF} & 0.0621 & 0.0338 & 0.0735 & 0.0705 & 0.0397 & 0.0636 & 0.0751 \\ 
        \textbf{UpliftRec-MF} & \textbf{0.0715} & \textbf{0.0363} & \textbf{0.0847} & 0.0873 & 0.0446 & 0.0720 & 0.0930 \\ 
        \textbf{FM} & 0.0532 & 0.0267 & 0.0630 & 0.0875 & 0.0416 & 0.0836 & 0.0932 \\ 
        \textbf{UpliftRec-FM} & 0.0555 & 0.0271 & 0.0657 & \textbf{0.1027} & \textbf{0.0504} & \textbf{0.0984} & \textbf{0.1093}  \\ 
        \textbf{LightGCN} & 0.0614 & 0.0299 & 0.0727 & 0.0796 & 0.0374 & 0.0777 & 0.0848 \\ 
        \textbf{UpliftRec-LGN} & 0.0650 & 0.0314 & 0.0770 & 0.0851 & 0.0394 & 0.0968 & 0.0907 \\ 
        \bottomrule
    \end{tabular}
    }}
    \end{center}

\label{tab:backend_per}
\end{table*}

\subsubsection{\textbf{Ablation Analysis}} We conduct ablation studies to analyze the effect of inverse propensity weighting on three datasets. We use R@10 for accuracy and RUP@10 for serendipity to examine the necessity of this IPW structure. $\gamma$ is the exponent of the propensity, which controls the extent of debiasing. As shown in Figure~\ref{fig:abs}, results on accuracy and serendipity both peak at a non-zero $\gamma$ on the three datasets. When $\gamma$ equals 0, the IPW structure does not work and relegates the model to suboptimal performance. The increasing trend demonstrates the debiasing effects of IPW because it concerns more seldom exposed ratios when estimating ADRF. The downward trends after reaching the highest points are also reasonable, as an overly debiased method gives too much weight to samples with tiny propensities, which compromises the model's performance.

\subsubsection{\textbf{MTEF vs. ADRF}}
\begin{figure}[tb!]
\setlength{\abovecaptionskip}{0cm}
\setlength{\belowcaptionskip}{-0.45cm}
\centering
\includegraphics[scale=0.25]{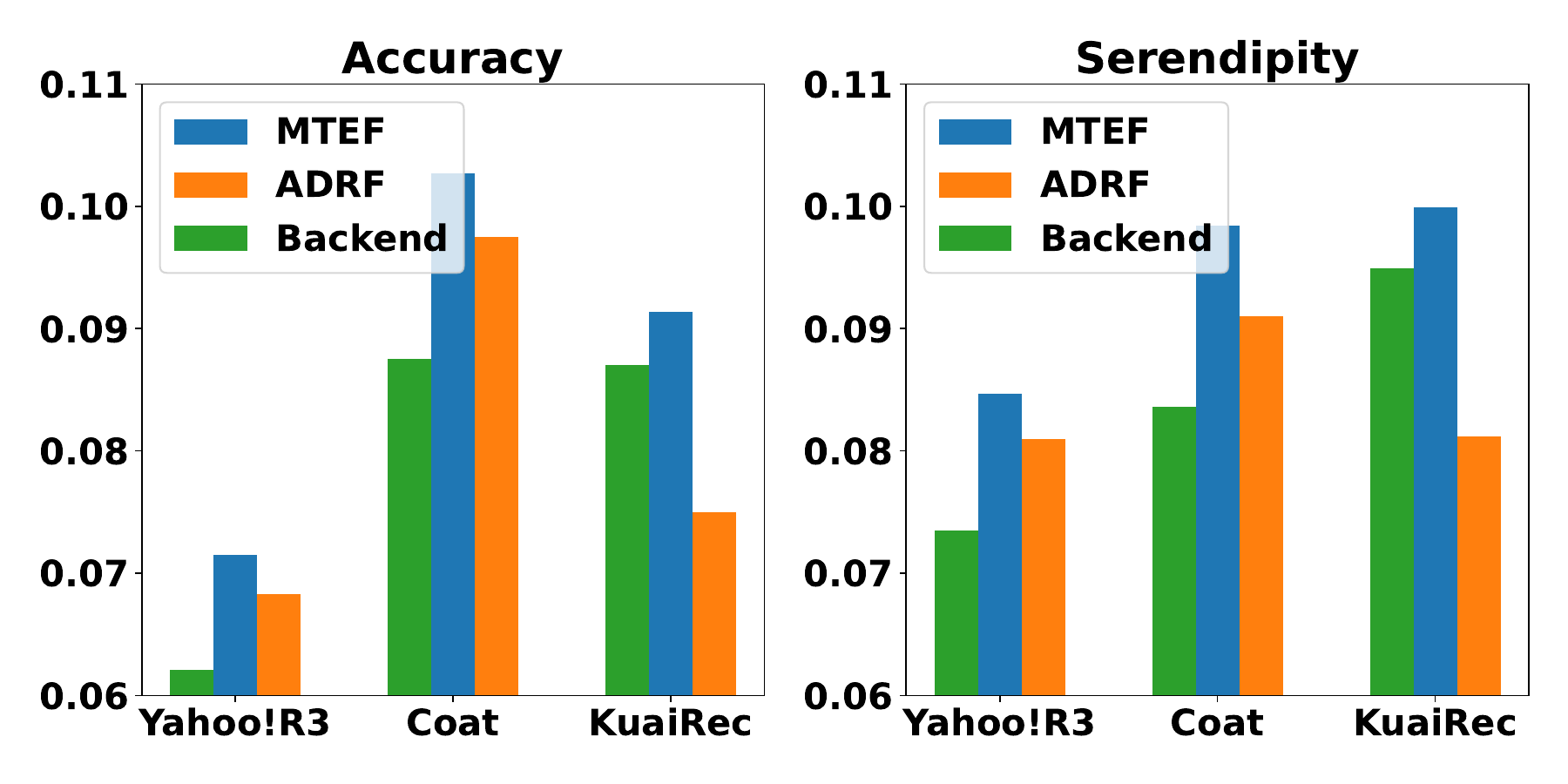}
\caption{Performance comparison of UpliftRec-METF and UpliftRec-ADRF on accuracy (left) and serendipity (right). We use R@10 as the metric for accuracy and use RUP@10 for Yahoo!R3 and RUE@10 for Coat and KuaiRec for serendipity.}
\label{fig:MTEFvsADRF}
\end{figure}

We compare the performances of UpliftRec-MTEF and UpliftRec-ADRF on three datasets (Figure \ref{fig:MTEFvsADRF}). UpliftRec-MTEF surpasses UpliftRec-ADRF across all datasets by leveraging marginal treatment effects to mitigate the high variance issue. However, UpliftRec-ADRF still achieves better performances on Yahoo!R3 and Coat compared to their backend models. The unstable performance of UpliftRec-ADRF can be attributed to the variance generated during ADRF estimation, especially when outcomes of certain category ratios are too scarce to make accurate estimations.

\subsubsection{\textbf{Backend Analysis}}

UpliftRec adopts a model-based approach. Backend models provide item embeddings for clustering and user embeddings for calculating similarity. These models also furnish backend scores to assist UpliftRec in ranking items within the same categories (UpliftRec-ADRF) and in re-ranking items with category-level uplift enhancements (UpliftRec-MTEF). To validate the versatility of UpliftRec-MTEF across various backend models, we manipulate the scores of different backend models while keeping the item and user embeddings unchanged. The outcomes are presented in Table \ref{tab:backend_per}, revealing substantial improvements in both accuracy and serendipity achieved by our method.

\subsubsection{\textbf{Case Study}} 
\begin{figure}[t!]
\setlength{\abovecaptionskip}{0cm}
\setlength{\belowcaptionskip}{-0.45cm}
    \includegraphics[width=1.0\linewidth]{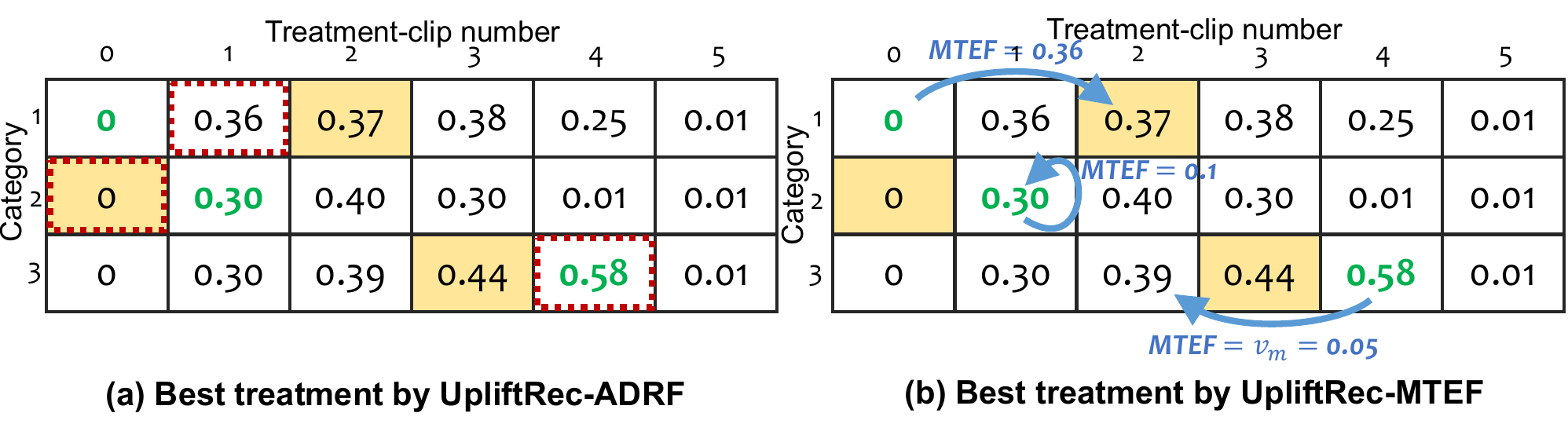}
    \caption{Visualization of the best treatment selection of UpliftRec-ADRF and UpliftRec-MTEF on Coat. The treatment provided by the backend model is highlighted in green, and the ground-truth treatment is in yellow-shaded boxes. The treatment calculated by ADRF is framed by red dotted lines in (a). The blue arrows point from the backend treatment to the treatment adjusted by MTEF in (b).}
    \label{fig:ADRF_vis}
\end{figure}
We present a case study on the Coat dataset to illustrate the workings of UpliftRec-ADRF and UpliftRec-MTEF. In Figure \ref{fig:ADRF_vis}, we visualize the application of these methods based on the estimated ADRF matrix of user-39. The ADRF matrix is displayed in a white and grey matrix format, where color intensity represents the magnitude of the ADRF value. With a matrix size of $3\times5$, corresponding to 3 clustered categories and a treatment-clip number $K$ of 5, null values for ADRF $v_a$ and MTEF $v_m$ are set at 0.01 and 0.05, respectively. Our objective is to predict the top-5 items for user-39, whose actual quantized category ratios $\Tilde{t}_g$—highlighted in yellow—amount to 2, 0, and 3 for each category. The backend model offers item scores and a recommendation list, denoted as $\Tilde{t}_0$, with quantized category ratios of 0, 1, and 4 (indicated in green in Figure \ref{fig:ADRF_vis}). The backend model, however, fails to identify user-39's latent interest, which is the first clustered category.

In UpliftRec-ADRF, we set the deviation acceptance $\epsilon$ to 2. Consequently, we compute the optimal ratios $\Tilde{t}_a$, framed by red dotted lines, which equate to 1, 0, and 4. Notably, ADRF successfully uncovers the hidden interest.

Meanwhile, in UpliftRec-MTEF, we use a step size $\Delta t$ of 1. The marginal treatment effects, which is calculated as $\mathrm{ADRF}(\Tilde{t}_0+\mathbf{1})-\mathrm{ADRF}(\Tilde{t}_0)$, yield [0.36, 0.1, 0.05]. The last term is padded with $v_m$ due to missing data. Substantial uplift is observed in the first category. The final ratios of UpliftRec-MTEF $\Tilde{t}_m$ are 2, 1, and 2. Among all these approaches, only UpliftRec-MTEF  successfully recommends item-240, the second-ranked item within the first clustered category,  showcasing its ability to to uncover user-39's latent interests.

\subsubsection{\textbf{Hyper-parameter Analysis}} 
UpliftRec leverages critical hyper-parameters, including $K_s$ for ADRF estimation, $K$ for treatment-clip binning, 
$C$ for category clustering, and $\lambda$ for the dividing ratio. Through experiments on Yahoo!R3 and Coat, we methodically assess how these parameters affect UpliftRec-MTEF's efficacy.

\begin{table*}[t]
\setlength{\abovecaptionskip}{0.15cm}
\setlength{\belowcaptionskip}{0cm}
\caption{Performance comparison between UpliftRec-MTEF and debiased methods with RCT data on the Yahoo!R3 and Coat. The best results are highlighted in bold while the best results besides AutoD-0.25 and InterD-0.25 are underlined.}
    \centering
    \setlength{\tabcolsep}{6mm}{
    \resizebox{\textwidth}{!}{
    \begin{tabular}{l|ccc|cccc}
    \toprule
        \textbf{Dataset} & \multicolumn{3}{c|}{\textbf{Yahoo!R3}} & \multicolumn{4}{c}{\textbf{Coat}} \\ 
        \textbf{Metric} & \textbf{R@10} & \textbf{N@10} & \textbf{RUP@10} & \textbf{R@10} & \textbf{N@10} & \textbf{RUE@10} & \textbf{RUP@10} \\ \midrule
        \textbf{MF} & 0.0621 & 0.0338 & 0.0735 & 0.0705 & 0.0397 & 0.0636 & 0.0751  \\ 
        \textbf{FM} & 0.0532 & 0.0267 & 0.0630 & 0.0875 & 0.0416 & 0.0836 & 0.0932  \\ 
        \textbf{UpliftRec-MTEF} & \underline{0.0715} & \underline{0.0363} & \underline{0.0847} & \underline{0.1027} & 0.0504 & 0.0984 & \underline{0.1093}  \\ 
        \textbf{AutoDebias-0.05} & 0.0587 & 0.0296 & 0.0695 & 0.0953 & \underline{0.0573} & \underline{0.1034} & 0.1015  \\ 
        \textbf{InterD-0.05} & 0.0603 & 0.0302 & 0.0714 & 0.0930 & 0.0565 & 0.0949 & 0.0991  \\ 
        \textbf{AutoDebias-0.25} & 0.0699 & 0.0358 & 0.0828 & \textbf{0.1089} & 0.0553 & \textbf{0.1034} & \textbf{0.1160}  \\ 
        \textbf{InterD-0.25} & \textbf{0.0746} & \textbf{0.0381} & \textbf{0.0884} & 0.1076 & \textbf{0.0598} & 0.0977 & 0.1146 \\ \bottomrule
    \end{tabular}
    }}

\label{tab:debiased_per}
\end{table*}

\begin{itemize}[leftmargin=*] 
    \begin{figure}[tb!]
    \setlength{\abovecaptionskip}{0cm}
    \setlength{\belowcaptionskip}{-0.4cm}
    \centering
    \includegraphics[scale=0.54]{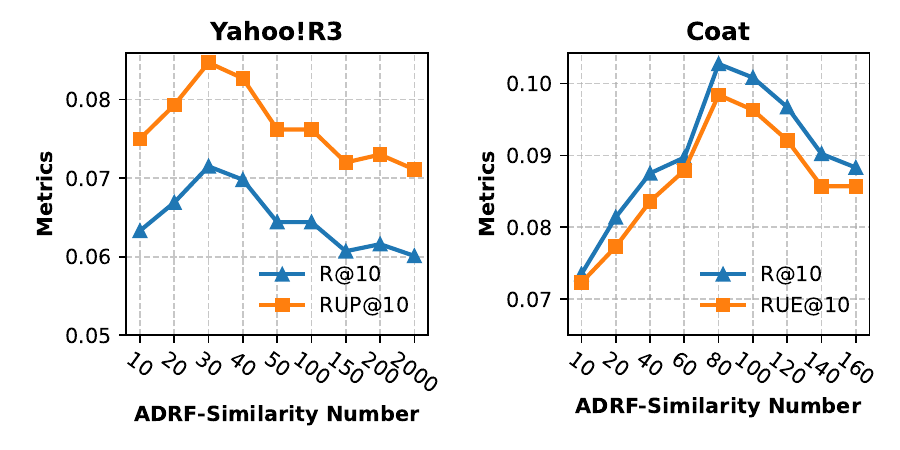}
    \caption{Performance of UpliftRec-METF on Yahoo!R3 (left) and Coat (right) under varying ADRF-similarity numbers.}
    \label{fig:ADRF-similarity}
    \end{figure}
    
    \item \textbf{The ADRF-similarity number $K_s$.} 
    UpliftRec employs the top $K_s$ similar samples for ADRF estimation. Setting $K_s$ too small can lead to challenges in accurately estimating ADRF due to high variance and insufficient data. Conversely, an excessively large $K_s$ may result in UpliftRec losing its personalization capability, leading to homogenous treatment outcomes. As illustrated in Figure \ref{fig:ADRF-similarity}, the performance of UpliftRec-MTEF peaks at $K_s=30$ for Yahoo!R3 and $K_s=80$ for Coat. This emphasizes the importance of choosing an appropriate value for $K_s$ to achieve optimal results.

    \begin{figure}[tb]
    \setlength{\abovecaptionskip}{0cm}
    \setlength{\belowcaptionskip}{-0.55cm}
    \centering
    \includegraphics[scale=0.54]{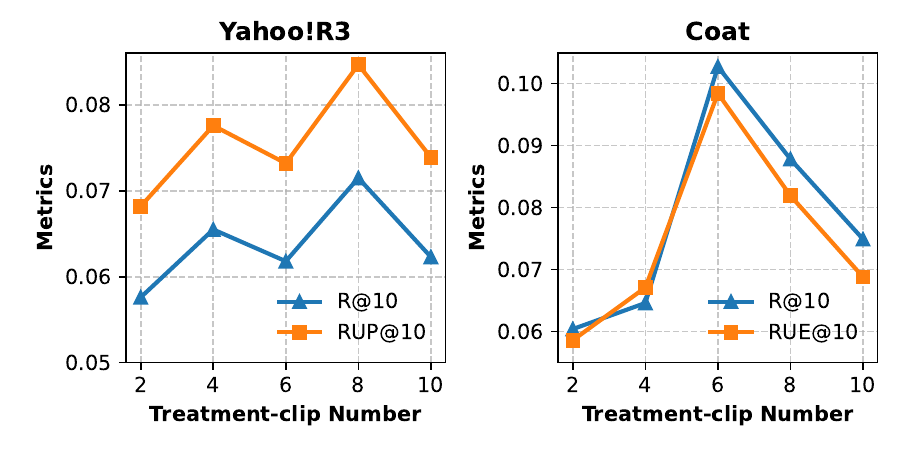}
    \caption{Performance of UpliftRec-METF on Yahoo!R3 (left) and Coat (right) under varying treatment-clip numbers.}
    \label{fig:treatment-clip}
    \end{figure}
    
    \item \textbf{The treatment-clip number $K$.} We discretize the continuous ratio space into $K$ quantized segments to address data scarcity. Figure \ref{fig:treatment-clip} illustrates the performance of UpliftRec-MTEF across different treatment-clip numbers $K$. Optimal $K$ values are found to be 8 for Yahoo!R3 and 6 for Coat. Choosing a small $K$ may fail to capture model complexities, and a large $K$ can cause data sparsity. The observed trends highlight the need for a proper $K$ value to balance these concerns.

    \begin{figure}[tb]
    \setlength{\abovecaptionskip}{0cm}
    \setlength{\belowcaptionskip}{-0.4cm}
    \centering
    \includegraphics[scale=0.52]{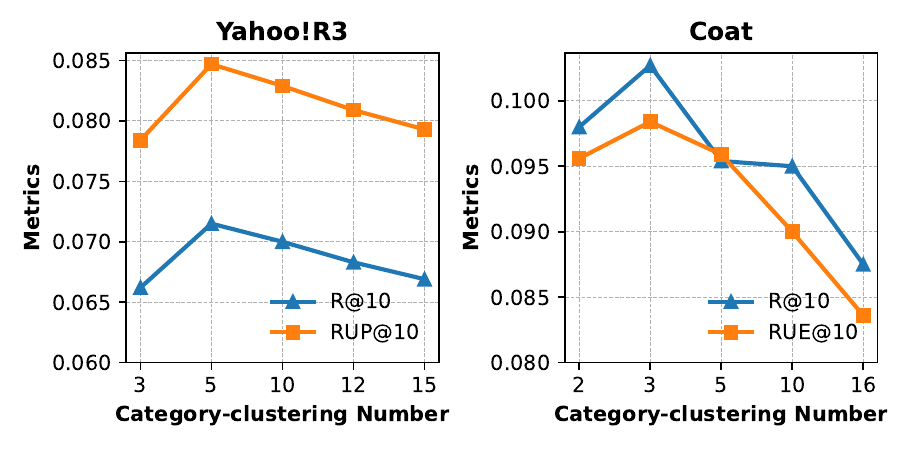}
    \caption{Performance of UpliftRec-METF on Yahoo!R3 (left) and Coat (right) under varying category-clustering numbers.}
    \label{fig:category-clustering}
    \end{figure}

    \item \textbf{The category-clustering number $C$.} 
    We perform item clustering, condensing items into $C$ new categories. Optimal clustering enhances matrix density, ensuring access to minority categories while enabling differentiation between each item category. Figure~\ref{fig:category-clustering} demonstrates that UpliftRec-MTEF's performance initially improves with an increasing number of categories $C$, reaching its peak at 5 categories for Yahoo!R3 and 3 categories for Coat, followed by a decline. Our findings indicate that proper clustering significantly boosts the performance of UpliftRec-MTEF.

    \begin{table}[t]
    \setlength{\abovecaptionskip}{0.1cm}
    \setlength{\belowcaptionskip}{0cm}
    \caption{Performance of UpliftRec-METF on Yahoo!R3 (left) and Coat (right) under different dividing ratios.}
        \centering
        \begin{center}
        \setlength{\tabcolsep}{2mm}{
        \resizebox{0.4\textwidth}{!}{
        \begin{tabular}{l|cc|ccc}
        \toprule
            \textbf{Dataset} & \multicolumn{2}{c|}{\textbf{Yahoo!R3}} & \multicolumn{3}{c}{\textbf{Coat}} \\ 
            \textbf{$\lambda$} & \textbf{R@10} & \textbf{RUP@10}  & \textbf{R@10} & \textbf{RUE@10} & \textbf{RUP@10}  \\ \midrule
            \textbf{0.25} & 0.0644  & 0.0763 & 0.0891 & 0.0836 &  0.0949 \\ 
            \textbf{0.5}  & \textbf{0.0715} & \textbf{0.0847} & \textbf{0.1027} & \textbf{0.0984} & \textbf{0.1093} \\ 
            \textbf{0.75} & 0.0675  & 0.0800 & 0.0831 & 0.0749  & 0.0885 \\ 
            \bottomrule
        \end{tabular}
        }
        }
        \end{center}
    \label{tab:lmd}
    \end{table}
    
    \item \textbf{The dividing ratio $\lambda$.} {We divide the training dataset into a history part and a treatment-outcome part using the ratio $\lambda$. An analysis was conducted by varying $\lambda$, and the corresponding results for Yahoo!R3 and Coat are presented in Table~\ref{tab:lmd}. We find that choosing $\lambda=0.5$ strikes a favorable balance between estimating user historical preferences and treatment effects.
}
    
\end{itemize}


\subsubsection{Comparison with Methods with Unbiased Data}

To better compare UpliftRec-MTEF with SOTA debiased methods, we compare it with two debiased methods employing a small proportion of RCT data in training in Yahoo!R3 and Coat, which are:
\begin{itemize}[leftmargin=*]
    \item \textbf{AutoDebias~\cite{chen2021autodebias}.} It is a strong debiased model trained with normal biased and RCT training data. It optimizes propensity scores and an imputation model over RCT training data. 
    \item \textbf{InterD~\cite{ding2022interpolative}.} 
    It is a distillation model that considers MF and AutoDebias as biased and debiased teachers, respectively, aiming to perform well in both biased and unbiased tests.
\end{itemize}
We split the original validation dataset into the unbiased training part and the current validation part. We set the proportion of the unbiased training part as 0.05 and 0.25 respectively of the entire unbiased dataset. Specifically, when the ratio is 0.05, we split the unbiased dataset into an unbiased training set (0.05), a validation set (0.45), and a testing set (0.5). When the ratio is 0.25, the proportions are 0.25, 0.25, and 0.5 respectively. We train the model with both the previously employed biased training data and the additional unbiased training set. We can see from Table \ref{tab:debiased_per} that the performance of UpliftRec-MTEF is somewhere between debiased-0.05 and debiased-0.25. To be specific, UpliftRec-MTEF outperforms both InterD-0.05 and AutoDebias-0.05 on accuracy and serendipity in Yahoo!R3, and have comparable results with them in Coat. When the ratio is raised to 0.25, InterD and AutoDebias act better than all the other methods, which demonstrates the strong ability of imputation-based debiased methods with unbiased training data. 

\section{Related Works}

\noindent$\bullet$ \noindent\textbf{User Interest Exploration.} Exploring user hidden great improves users' satisfaction. Existing works in this area can be categorized into three groups: debiased methods~\cite{boratto2021connecting,wang2019doubly,saito2020unbiased,chen2021autodebias,ding2022interpolative}, diversity-oriented methods~\cite{carbonell1998use,sha2016framework,stamenkovic2022choosing}, and bandit methods~\cite{li2010contextual,li2016collaborative,song2022show}.

\begin{itemize}[leftmargin=*]
    \item Debiased methods aim to mitigate biases~\cite{chen2023bias} in recommender systems, addressing issues such as selection bias~\cite{schnabel2016recommendations}, popularity bias~\cite{zhang2021causal, zhou2023adaptive}, and exposure bias~\cite{hu2008collaborative}. Recently, new debiased methods have emerged, exploring additional aspects including fairness bias~\cite{tang2023fairness}, duration bias~\cite{zhao2023uncovering}, and outlier bias~\cite{sarvi2023impact}.
    Many methods have been proposed to mitigate bias, such as regularization~\cite{boratto2021connecting}, data imputation~\cite{wang2019doubly}, causal inference~\cite{sun2022counterfactual, pearl2016causal}. Sam-reg~\cite{boratto2021connecting} introduces regularization to mitigate the biased correlation between user-item relevance and item popularity. Doubly robust~\cite{wang2019doubly} trains a data imputation network to estimates the effect of missing data. IPS~\cite{saito2020unbiased}, as a widely used causal inference method, is designed to alleviate different effects of biased feedback. AutoDebias~\cite{chen2021autodebias} further introduces a general debiased framework capable of combating multiple biases and their combinations using meta-learning. InterD~\cite{ding2022interpolative} adopts an interpolative framework and leverages other methods as teachers to obtain a model that is both robust on biased and debiased datasets. However, they lack the ability to actively discover new interests. 
    
    \item Diversity-oriented methods aim to find users' new interests by increasing the categories of recommended items. They recommend diversified items mainly in two ways: post-processing methods~\cite{carbonell1998use,sha2016framework} and diverse-objective methods~\cite{stamenkovic2022choosing}. Post-processing methods are applied during the inference stage of the recommender systems to improve the coverage of all categories without changing the representation of users and items. On the other hand, diverse-objective methods optimize a diversity-oriented objective while training the models. For example,~\cite{stamenkovic2022choosing} adds a reward for diversity when training the reinforcement learning framework. 
    However, both diversity-oriented methods focus on covering all categories, irrespective of user relevance. This compromises recommender system performance as low-CTR categories occupy excessive exposure resources.
    
    \item Bandit methods consider the potential rewards of enhanced exploration-exploitation trade-offs. They estimate each arm's mean reward and variance (item or item group) by iteratively collecting new user feedback. Bandit methods explore according to the estimation of the mean and variance of each arm~\cite{li2010contextual}. COFIBA~\cite{li2016collaborative} considers collaborative effects and clusters users and items based on traditional bandit algorithms. HCB~\cite{song2022show} proposes a hierarchical contextual bandit algorithm to reduce the computation cost when the number of arms is large. GNB~\cite{qi2023graph} estimates the user graphs to preserve the pair-wise user correlations and utilize individual GNN-based models to achieve the adaptive exploration. However, Bandit methods still have a high chance to explore dissatisfying items owing to the high variance of these items. Moreover, the absence of list-wise optimization in these methods results in suboptimal recommendation lists.
\end{itemize}


\noindent$\bullet$ \noindent\textbf{Uplift Modeling.} Uplift modeling~\cite{zhang2021unified, gutierrez2017causal} is employed to estimate the incremental effects of treatments (e.g., coupon distribution~\cite{ai2022lbcf}). It tackles a counterfactual problem since we cannot directly observe the effects of intervention or non-intervention. Several methods have been proposed to address this challenge. LBCF~\cite{ai2022lbcf}, uses causal forests to estimate the conditional average treatment effect (CATE), and proposes a budget constraint optimization algorithm, while the treatment is binary. $MDP^2$ Forest~\cite{yu2022mdp2}, employs a Forest Ensemble approach where each tree in the forest estimates a greedy treatment (\ie category distribution). Nevertheless, current methods do not guarantee optimality and additional random experiments are needed to obtain the data, which causes harm to user experience and company revenue. 

\noindent$\bullet$ \noindent\textbf{Causal Recommendation.} In recent years, causal recommendation methods~\cite{zhang2021causal,mehrotra2020inferring,sato2020unbiased,christakopoulou2020deconfounding,zhang2023debiasing} have been continuously introduced, such as works on counterfactual learning and on confounding effects. For counterfactual learning,~\cite{mehrotra2020inferring} utilize a Bayesian framework to qualify the treatment effects to counterfactual data from unobserved treatments. ~\cite{zhang2023debiasing} leverages proxy variables to infer the unmeasured confounder and users’ feedback. For de-confounding methods, DLCE~\cite{sato2020unbiased} regards the features of users and items as confounders and re-weight the training samples to learn treatment effects. Another method~\cite{christakopoulou2020deconfounding} considers the response rate as the confounder and employs IPW to debias. 

\section{Conclusion}
In this work, we approach top-$N$ recommendation from a causal perspective, using the category exposure ratio as the treatment and the CTR of each category as the outcome. We tackle the challenges of expensive data collection, biased observation, and data sparsity by generating an augmented dataset for treatment effect estimation, employing an IPW method to address the backdoor path, and discretizing the treatment. Our proposed Uplift model-based Recommender (UpliftRec) framework estimates personalized ADRF to explore user interests effectively. By optimizing potential outcomes based on ADRF, we determine the best treatment for each user and adjust backend scores using MTEF derived from ADRF.We conduct extensive experiments to validate the effectiveness of our framework.
This work introduces a new field of estimating uplift effects in recommendations. In the future, we plan to extend our approach to session-based models, such as reinforcement learning. Furthermore, we will explore different actions as treatments to apply our method to various scenarios.

\bibliographystyle{ACM-Reference-Format}
\bibliography{main}


\end{document}